\documentclass[conference]{IEEEtran}
\IEEEoverridecommandlockouts
\usepackage{cite}
\usepackage{amsmath,amssymb,amsfonts}
\usepackage{algorithmic}
\usepackage{graphicx}
\usepackage{textcomp}
\usepackage{xcolor}
\usepackage{subcaption}
\usepackage{url}
\usepackage{framed}
\def\BibTeX{{\rm B\kern-.05em{\sc i\kern-.025em b}\kern-.08em
    T\kern-.1667em\lower.7ex\hbox{E}\kern-.125emX}}
    
\newcommand{\featurelm}{\textit{LM}}
\newcommand{\featuredtov}{\textit{Doc2Vec}}
\newcommand{\featuretfidf}{\textit{TF*IDF}}
\newcommand{\featurecount}{\textit{COUNT}}

\newcommand{\rom}[1]{\uppercase\expandafter{\romannumeral #1\relax}}

\usepackage[numbers]{natbib}

\begin{document}

\title{Semantic Text Analysis for Detection of Compromised Accounts on Social Networks}

\author{\IEEEauthorblockN{Dominic Seyler, Lunan Li, and ChengXiang Zhai}
\IEEEauthorblockA{\textit{Department of Computer Science} \\
\textit{University of Illinois at Urbana-Champaign}\\
\{dseyler2, lunanli3, czhai\} @illinois.edu}
}

\maketitle

\begin{abstract}
Compromised accounts on social networks are regular user accounts that have been taken over by an entity with malicious intent. Since the adversary exploits the already established trust of a compromised account, it is crucial to detect these accounts to limit the damage they can cause. We propose a novel general framework for semantic analysis of text messages coming out from an account to detect compromised accounts.  Our framework is built on the observation that normal users will use language that is measurably different from the language that an adversary would use when the account is compromised. We propose to use the difference of language models of users and adversaries to define novel interpretable semantic features for measuring semantic incoherence in a message stream. We study the effectiveness of the proposed semantic features using a Twitter data set. Evaluation results show that the proposed framework is effective for discovering compromised accounts on social networks and a KL-divergence-based language model feature works best.
\end{abstract}


\section{Introduction}\label{sec:introduction}
The advent of social media has borne great opportunities and benefits, but also dangers and risk. One problem that has become more prominent is the spread of misinformation on social networks. In order to spread misinformation successfully, perpetrators mainly rely on social/spam bots \cite{cresci2019better,DBLP:conf/ccs/GrierTPZ10,DBLP:journals/eswa/Martinez-RomoA13} or compromised accounts \cite{DBLP:conf/ccs/ThomasLGP14,DBLP:conf/sac/ZangerleS14}. 
Account compromising has become a major issue for public entities and regular users alike.
In 2013, over a quarter million accounts on Twitter were compromised \cite{cnn-twitter} and despite massive efforts to contain account hacking, it is still an issue today \cite{olmstead2017americans}.
For affected users a compromised account can be an embarrassing experience. As a result, 21\% of users that fall victim to an account hack abandon the social media platform~\cite{DBLP:conf/ccs/ThomasLGP14}. 

Compromised accounts are legitimate accounts that a malicious entity takes control over, with the intention of gaining financial profit \cite{DBLP:conf/ccs/ThomasLGP14} or spreading misinformation \cite{DBLP:conf/ndss/EgeleSKV13}. These accounts are especially interesting for attackers, as they can exploit the trust network that the user has established \cite{DBLP:journals/tifs/RuanWWJ16}. Since an account takeover can take up to five days, with 60\% of the takeovers lasting an entire day \cite{DBLP:conf/ccs/ThomasLGP14}, attackers are given ample time to reach their goal.
Finding these hijacked accounts is challenging, since they exhibit traits similar to regular accounts \cite{DBLP:journals/tifs/RuanWWJ16}. Only after analyzing the changes in the account's behavior, patterns can be identified that expose the account as being compromised and therefore specific methodology is required.

Existing work on detection of compromised accounts has mostly relied on anomalies of user profiles, but there is a great opportunity to leverage semantic analysis of an account's content, since the intent of compromising a social media account is to inject ``abnormal" content~\cite{karimiend2018e2e,vandam2018cadet,vandam2019you}. Thus, analysis of semantic coherence of text content can be a general strategy for detecting compromised accounts. While some existing work has attempted to leverage text content also, 
the textual features used are simplistic or non-interpretable~\cite{DBLP:conf/ndss/EgeleSKV13,vandam2017understanding,karimiend2018e2e}. For example, bag-of-word features are inadequate for capturing semantic variations in language, while embedding-based approaches are not interpretable, which is important if such a method is to be used to guide any real-world actions on the account. In this paper, we propose to perform deeper semantic analysis using a solid probabilistic language model framework to directly measure the semantic incoherence in text content, leading to highly interpretable semantic features for detecting compromised accounts. Such features can be used in any supervised learning framework to improve detection accuracy and improve explainability.

Our key observation is that a regular user's textual output will differ from an attacker's textual output. We thus propose a general framework for detecting compromised accounts based on semantic analysis of the incoherence in the text stream. This is complementary with the existing work in the sense that our framework can lead to highly interpretable novel features that can be added to any existing machine learning-based detection method to improve its accuracy and explainability.  As a specific implementation of the framework, we model the user's and attacker's language as two smoothed multinomial probability distributions that are estimated using the textual output of the user and attacker, respectively. We use a similarity measure between probability distributions as indicators that an account is being compromised. Even though we do not know the start and the end of an account takeover, our method leverages the fact that the average difference for random account begin and end dates will be higher for compromised accounts, compared to benign accounts.

Evaluation of such a detection task is challenging due to the inevitable concern of user privacy, making it nearly impossible to have a publicly available real world data set. Following other work in this domain~\cite{gupta2013modeling,DBLP:journals/tifs/RuanWWJ16,DBLP:conf/enic/TrangJR15,vandam2019you}, we propose a simulation-based evaluation method and use such a method to study the effectiveness of the proposed semantic analysis method. Specifically, we formalize this problem in a threat model and utilize this model to simulate account takeovers in our dataset, to compare an implementation of the proposed framework\footnote{The code is available at: \url{https://github.com/dom-s/comp-account-detect}.} to other approaches. 

Our evaluation results show that the proposed incoherence-based features are highly effective for detecting compromised accounts and can be combined with other features to improve detection accuracy and enhance explainability. 
We further show that when the proposed method is trained on simulated data, 
it can detect non-artificially compromised accounts from real world data set. 
Since training on simulated data requires no human effort, the proposed method can be immediately adopted by a social media company to  potentially enhance their compromised account detector.

\section{Related Work}
\label{sec:related-work}

It is common practice in compromised account detection to build profiles based on user behavior and look for anomalies within them. Egele et al.~\cite{DBLP:conf/ndss/EgeleSKV13} learns behavioral profiles of users and looks for statistical anomalies in features based on temporal, source, text, topic, user, and URL information. Ruan et al.~\cite{DBLP:journals/tifs/RuanWWJ16} finds anomalies in the variance of a user's click behavior. Viswanath et al.~\cite{DBLP:conf/uss/ViswanathBCGGKM14} applies principal component analysis to a user's Facebook likes to find abnormal behavior. Vandam et al.~\cite{vandam2017understanding} studies certain account characteristics, such as number of hashtags or number of mentions in tweets, which are used as features in a classification framework. Karimi et al.~\cite{karimiend2018e2e} uses Long Short-Term Memory networks to capture the temporal dependencies within user accounts to learn distinguishing temporal abnormalities. \citet{vandam2018cadet} uses an unsupervised learning framework, where multiple views on a user profile (i.e., term, source, time and place) are encoded separately and then mapped into a joint space. This joint representation is then used to retrieve a ranking of compromised accounts. Building on this work, \citet{vandam2019you} uses and encoder-decoder framework to build low-dimensional feature vectors for users and tweets. The residual errors form both encoders are used in a supervised setting to predict compromised accounts. 

When it comes to textual information, current methods either do not leverage it at all \cite{DBLP:conf/ndss/EgeleSKV13,DBLP:journals/tifs/RuanWWJ16} our the textual features are superficial. For instance, text is only used to detect language changes (e.g., from English to French) \cite{DBLP:conf/ndss/EgeleSKV13} or topics are derived plainly from hashtags \cite{DBLP:conf/ndss/EgeleSKV13,vandam2017understanding}. It is also common to simply use bag-of-word features \cite{vandam2017understanding} or neural embeddings \cite{karimiend2018e2e,vandam2018cadet,vandam2019you}. Thus, all methods can benefit of a deeper analysis of the semantic incoherence of text.

\section{Semantic Incoherence Framework}
\label{sec:appraoch}

\subsection{Threat Model}
\label{sec:threat-model}

The adversary's goal is to inject textual output into a benign account in order to mask the output's origin and leverage the user's influence network.  In our setting, it is irrelevant how the adversary obtained access to the account. 
To make our method as general as possible, we assume that no account details are observed, meaning that we ignore the friendship network, profile details, etc. To further cater to generality, we don't make any assumptions about the text, except that it was written by a different author. In the case of an attack, we assume that at least one message is injected by the adversary and that at least one benign message is observed. For all accounts we assume that they existed for more than one day.

\begin{figure}[t]
\centering
\includegraphics[width=0.95\linewidth]{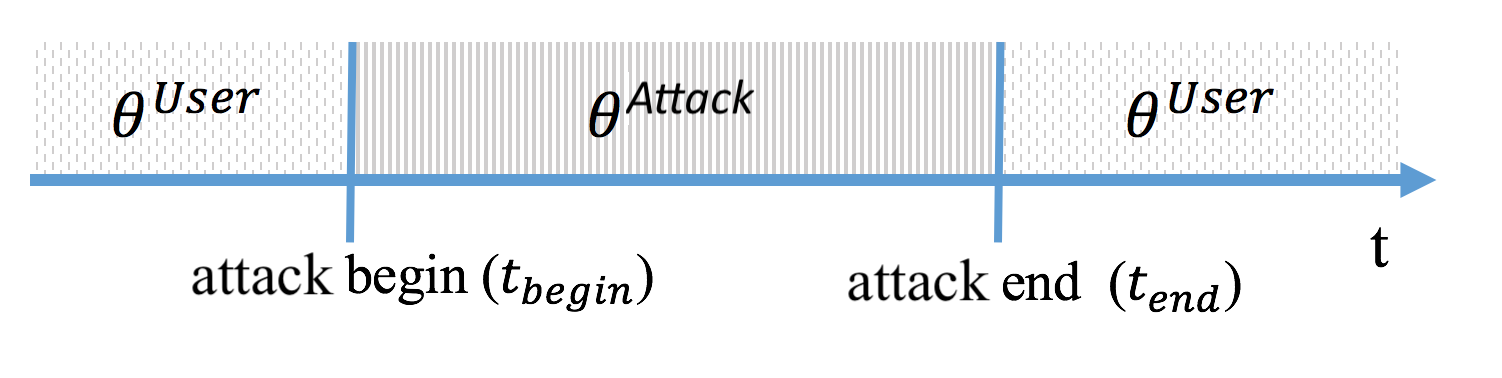}
\caption{A tweet stream divided according to begin ($t_{begin}$) and end ($t_{end}$) of an attack. A language model is learned for regular tweets ($\theta^{User}$) and compromised tweets ($\theta^{Attack}$). }
\label{fig:lms}
\vspace{+0.3cm}
\end{figure}

\begin{figure}[t]
\centering
\includegraphics[width=1\linewidth]{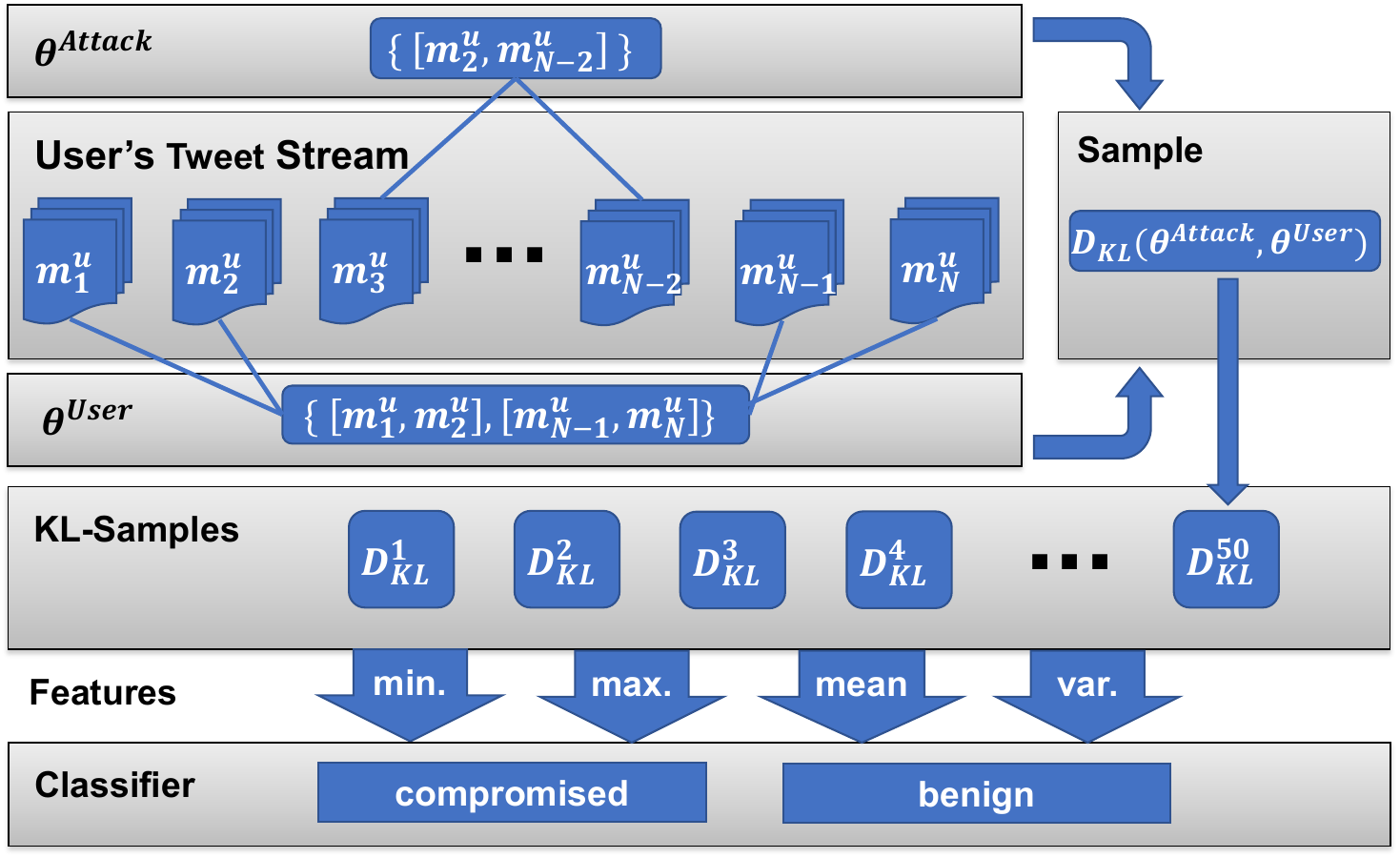}
\caption{Overview of model.}
\label{fig:approach-overview}
\end{figure}

\subsection{General Framework}

We now describe the proposed framework for identifying compromised accounts. In accordance with our threat model, the framework is based on the assumption that an adversary's textual output will deviate from a regular user's textual output.  Let $\mathcal{U}$ be the set of all users $u$. Further, let
$m_t^u$  be a message $m$ from user $\mathit{u}$ at time $t \in \{1, 2, ..., N\}$. Our goal is to find all compromised user accounts  $\mathcal{U}^{comp} \subset \mathcal{U}$.
To capture the discrepancy between language usage, we propose to divide the tweet space of a user into two non-overlapping sets $M^{User}$ and $M^{Attack}$. We randomly assign two timepoints: $t_{start}$ signals the start of the attack and $t_{end}$ signals its end. All $m_t^u$ with $t \in [t_{start}, t_{end}]$ make up $M^{Attack}$, whereas all $m_t^u$ with $t \in [t_1, t_{start-1}] \cup t \in [t_{end+1}, t_{N}]$ make up $M^{User}$. We can measure the difference between $M^{User}$ and $M^{Attack}$ using any similarity measure of our choice. This procedure can be repeated multiple times for different values of $t_{start}$ and $t_{end}$. $t$ can be of different granularity, where the minimum is per-message and the maximum can be chosen at will. The more often the procedure is repeated for a certain user, the higher is the sampling rate, which makes for better approximations of the true difference between $M^{User}$ and $M^{Attack}$ (we study the optimal sample rate in Section \ref{sec:experiments}). Thus, this strategy provides a flexible trade-off between accuracy and efficiency. The similarity measures can then be employed as features in the downstream task. 

\subsection{Instantiation with Language Modeling}

We describe our practical instantiation of the framework using language modeling and supervised learning. We create a classifier that distinguishes compromised from benign user accounts based on a feature set, derived from our framework. We measure this as the difference between two word probability distributions of a user and an attacker. We select KL-divergence \cite{kullback1951} as our method of choice to compare the difference of probability distributions. We assume that when a user writes a message she draws words from a probabilistic distribution that is significantly different from the distribution an attacker draws words from. Let $\theta^{User}$ and $\theta^{Attack}$ be two word probability distributions (i.e., language models) for the user and attacker. We need to select two time points $t_{start}$ and $t_{end}$ that mark beginning and end of the attack (Figure \ref{fig:lms}). As described by our framework, all messages $m_t^u$ that fall within the time interval $[t_{start}, t_{end}]$ contribute to $\theta^{Attack}$, whereas $m_t^u \notin [t_{start}, t_{end}]$ contribute to $\theta^{User}$. 

Figure \ref{fig:approach-overview} presents an overview of our model. For a particular user, we sample her tweet stream for random $t_{start}$, $t_{end}$ pairs. As seen in the figure, the algorithm can, for example, select $t_3$ and $t_{N-2}$ as $t_{start}$ and $t_{end}$, respectively. Thus, all tweets that fall between $\{[m^u_1, m^u_2], [m^u_{N-1}, m^u_N]\}$ contribute to $\theta^{User}$. All tweets that fall in $[m^u_3, m^u_{N-2}]$ contribute to $\theta^{Attack}$. Then, the KL-divergence for these specific $\theta^{User}$ and $\theta^{Attack}$ contributes one sample for this user ($D_{KL}^{50}$ in the figure). This process is repeated for different samples (e.g., we used 50 samples in our experiments), where each time $t_{start}$ and $t_{end}$ are selected at random, with constraint $t_{start} < t_{end}$. 
Sample rates are selected empirically, depending on the best classification performance on a development set.
We select the maximum, minimum, mean and variance of the sampled KL-divergence scores. These features are then combined in a Support Vector Machine (SVM) model, which learns the optimal weighting for each feature based on the training data. Naturally, other classifiers can also be used, but exploration of different classifiers is out of the scope of this paper.

\subsection{Language Modeling Details}

We try to estimate the joint probability of $P(w_1, w_2, w_3, ..., w_N)$ for all words $w_1, ..., w_N$ in the text. According to the chain-rule, this is equivalent to computing $P(w_1)P(w_2|w_1)P(w_3|w_1,w_2)... P(w_N|w_1, ..., w_{N-1})$. Because of the combinational explosion of word sequences and the extensive amount of data needed to estimate such a model, it is common to use n-gram language models. N-gram language models are based on the Markov assumption that a word $w_1$ only depends on $n$ previous words ($P(w_k|w_1, ..., w_{k-1}) \approx P(w_k|w_{k-n+1}, ..., w_{k-1})$). The simplest and computationally least expensive case is the uni-gram. Here, the Markov assumption is that a word $w_k$ is independent of any previous word, i.e., $P(w_k|w_1, ..., w_{k-1}) \approx P(w_k)$. All $P(w_k)$ for $k \in 1, 2, ..., N$ make up a language model $\theta$, which is a multinomial probability distribution, where words in the document are events in the probability space. The parameters for the language models $\theta$ are estimated using maximum-likelihood. Maximizing the likelihood of the uni-gram language model is equivalent to counting the number of occurrences of $w_k$ and dividing by the total word count ($P(w_k) = \frac{c(w_k)}{N}$, where $c(w_k)$ is the word count of $w_k$). This distance between two probability distributions can be estimated using Kullback-Leibler-divergence \cite{kullback1951}. In the discrete case, for two multinomial probability distributions $P$ and $Q$ the KL-divergence is given as \[D_{KL}(P,Q) = \sum_i P(i) log(\frac{P(i)}{Q(i)}).\] It can be observed that $D_{KL}(P,Q) \neq D_{KL}(Q,P)$. However, it is common practice to still think of $D_{KL}$ as a distance measure between two probability distributions \cite{zhai2016text}. An issue with $D_{KL}$ is that the sum runs over $i$, which is the event space of $P$ and $Q$. Thus, it requires the event space to be equivalent for both distributions. In our case $\theta$ is a language model.  Now, let $v(\theta)$ denote the vocabulary set of $\theta$. As a result of maximum-likelihood estimation, in most cases $v(\theta^{User}) \neq v(\theta^{Attack})$. Thus, we have to smooth the probability distributions such that $v(\theta^{User}) = v(\theta^{Attack})$, which is required in order to calculate $D_{KL}$. To achieve this, we define $v(\theta) := v(\theta^{User}) \cup v(\theta^{Attack})$. Then, we set $v(\theta^{User}) = v(\theta)$ and $v(\theta^{Attack}) = v(\theta)$ and estimate $\theta^{User}$ and $\theta^{Attack}$ using the Laplace estimate.

\section{Experimental Design}
\label{sec:experimental-design}

We turn the compromised account detection task into a binary classification problem, where the goal is to decide whether a user account is compromised or not. We therefore learn a classification function, which returns 1 if an account is compromised and 0 otherwise.

We set up our experiments to answer the following research questions: We first perform a feasibility analysis to \textbf{(\rom{1})} study to what extend a KL-divergence measure can detect incoherence. 
The second part of our feasibility analysis finds proof that \textbf{(\rom{2})} the average KL-divergence can be estimated by randomly sampling a certain number of points with different begin/end dates. 
Following the feasibility study, we show how effective the proposed language model based method is in detecting compromised accounts in a simulated environment.
We show \textbf{(\rom{3})} how the proposed features compare to general state-of-the-art text classification features and how to combine them;
\textbf{(\rom{4})} how our proposed features perform in comparison to other compromised account detection methods and how we can further improve performance by combining their feature spaces;
\textbf{(\rom{5})} how effective is our method on a real (non-simulated) dataset by performing a qualitative, manual investigation. 
We start by introducing our dataset.

\subsection{Dataset}
\label{ssec:dataset-creation}

Previous methods on compromised account detection either do not publish their datasets \cite{DBLP:conf/ndss/EgeleSKV13,vandam2017understanding,DBLP:conf/uss/ViswanathBCGGKM14} or the original text data is not fully recoverable due to restrictions of the social media platform or the deletion of user data \cite{karimiend2018e2e}. Therefore, meaningful evaluation is especially challenging in our setting. Since no dataset exists, we opted to follow existing practise~\cite{gupta2013modeling,DBLP:journals/tifs/RuanWWJ16,DBLP:conf/enic/TrangJR15,vandam2019you} and use a simulation method. We perform a simulation of account attacks, which allows us to compare different methods quantitatively to study their effectiveness. Any bias introduced by the simulation is unlikely to affect our conclusions as it is orthogonal to the methods that we study and we argue that such an evaluation strategy is adequate to perform a fair comparison of different methods.

For simulation, we leverage a large Twitter corpus from Yang and Leskovec \cite{DBLP:conf/wsdm/YangL11}. The dataset contains continuous tweet streams of users. Relationships between users, etc. are unknown. Finding compromised accounts manually within a dataset of millions of tweets is clearly infeasible.
Simulating account hijackings enables us to (cheaply) create a gold standard while having full control over the amount of accounts compromised and begin/end of account takeovers. For  simulation, we follow our threat model introduced in Section \ref{sec:threat-model}, where the adversary's goal is to inject messages into regular accounts. Since we make no assumptions about the textual content, we follow a similar methodology to \citet{vandam2019you} and \citet{DBLP:conf/enic/TrangJR15}, and replace part of the consecutive tweet stream of an account with the same amount of consecutive tweets from another random user account. Since our algorithm has no knowledge about the begin and end of an attack, we choose these at random, meanwhile ensuring that only a certain predefined fraction of the tweets is compromised. This allows us to test the effectiveness of our method in different scenarios, where different percentages of tweets are compromised within an account. To get meaningful estimates for the language models $\theta^{User}$ and $\theta^{Attack}$, we select the 100,000 users with most tweets in our data. 
According to our threat model, we retain users with more than one day of coverage, resulting in 99,912 users and 129,442,760 tweets.

\subsection{Feasibility Studies}

In our feasibility studies we use a subset of the data by selecting 495 users at random, where each compromised account contains 50\% compromised tweets. For each user, we put all tweets into daily buckets and calculate the KL-divergence for all possible combinations of $t_{begin}$ and $t_{end}$. The reason this cannot be done for the whole dataset is simply because of computational cost. We therefore operate on this subset to investigate whether our method is feasible and can be approximated to avoid increased computational effort.

\subsection{Quantitative Evaluation}

Our quantitative evaluation aims to show the performance of the algorithm using standard metrics for performance measure. We measure classification Accuracy; Precision, Recall and $F_1$-score for the class representing the predictions of compromised accounts (1-labels). In addition, we show effectiveness of our method for different levels of difficulty. We experiment with various settings for the percentage of compromised tweets in compromised accounts with random begin and end dates of account take over. More concretely, we select 50\%, 25\%, 10\% and 5\% of tweets to be compromised, each representing a more difficult scenario. As these ratios would not be strictly separated in a real-world scenario, we further experiment with random (``RND'') ratios, drawn uniformly from [5\%,50\%].
In our experiments the probability of an account being compromised is set to $0.5$ to obtain a balanced dataset\footnote{In reality the amount of compromised accounts is much lower then 50\%. However, it is common in supervised frameworks for compromised account detection to balance datasets to learn a better discriminative function \cite{karimiend2018e2e,vandam2019you}.}. Furthermore, we employ ten-fold cross validation.

\subsection{Baselines}

To study the effectiveness of our proposed features based on language modeling (\featurelm{}), we compare them to general text classification features (i.e., word-based and Doc2Vec \cite{le2014distributed}) and task-specific features from two methods in compromised account detection, namely COMPA \cite{DBLP:conf/ndss/EgeleSKV13} and VanDam \cite{vandam2017understanding}. 
We combine them in a Support Vector Machine (SVM) framework, which has been shown to be an excellent predictor for text classification. However, our approach is general and can be applied to any classifier.

\noindent \textbf{SVM}. We use SVM with linear kernels for all models. Features are standardized by removing the mean and scaling to unit variance. Ten-fold cross validation is performed for all models to ensure there is minimal bias in selecting the training/testing split of the data. All posts of a single user are concatenated as one document. The labels for each document were chosen as 1 if an account is compromised and 0 otherwise. 

\noindent \textbf{Word-based Features}. For word-based feature representation and feature selection we follow the methodology of Wang et al. \cite{wang2017study}. We choose count based (\featurecount) and TF*IDF based (\featuretfidf) representations of words and experiment with two supervised feature selection strategies, namely chi-square and mutual information. For dictionary creation we keep 100,000 uni-grams, which appear in no less than 20 and no more than 9991 (=10\%) of documents, to remove very rare and very common words.

\noindent \textbf{Document-based features (\featuredtov{})}. We leverage the method proposed in Le and Mikolov \cite{le2014distributed} to learn low-dimensional vector representations for documents. As recommended, we choose a vocabulary size of 2M tokens \cite{pennington2014glove}, 400-dimensional  document vectors and a window size of 5  \cite{le2014distributed}. We make use of the document vectors as features in our SVM classification framework.

\noindent \textbf{COMPA} \cite{DBLP:conf/ndss/EgeleSKV13}. This method builds behavioral user profiles and detects compromised accounts by finding statistical anomalies in the features. We implement features that can be derived from our dataset: time (hour of day), message text (language), message topic, links in messages and direct user interaction. This method is used for stream processing; once a user profile is build, the method checks a new message against the existing user profile. For each feature the message will be assigned an anomaly score, which reflects how different this message is from the previously observed user profile. Since our method classifies accounts as a whole, we simulate this process for each message within an account (in temporal order of the posting of the message). We then aggregate the anomaly scores, per feature, as the mean of the scores of all the account's messages.

\noindent \textbf{VanDam} \cite{vandam2017understanding}. This work investigated the distributional properties of different features on compromised accounts and uses them in a classification framework. We implement the following features from this method: hashtags, mentions, URLs, retweets and sentiment. The authors use these features to classify single messages as compromised. In our setting, we aggregate each feature's counts over all messages within an account and use the mean of the counts as features.

\subsection{Qualitative Evaluation}

Our qualitative evaluation aims to show how well compromised accounts can be detected in the original dataset. Therefore, we manually evaluate the accounts that have the highest probability according to our classifier. Here, a major change is that we only inject messages into the \textit{training} dataset, whereas the testing set is left untouched. We randomly select 70$\%$ of users for training and 30$\%$ of users for testing, with 25$\%$ of tweets compromised. Since there is no ground truth in the testing dataset, we define four specific metrics to evaluate whether an account is considered compromised. If two or more of the following metrics are met, an account is considered as being compromised: In certain time periods the account (1) has a sharp topic change, (2) has a specific posting frequency change, (3) has a specific language change (4) posts repeated tweets.

\section{Experiment Results}
\label{sec:experiments}

\begin{figure*}
 	\centering
 	\begin{subfigure}{0.24\textwidth}
 		\includegraphics[width=0.9\linewidth,keepaspectratio]{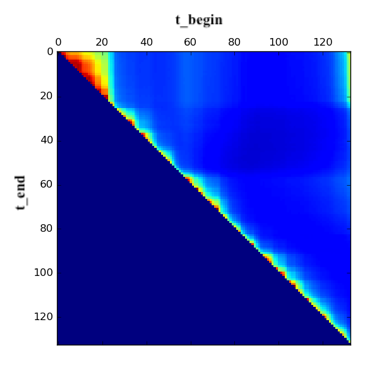}
 		\vspace{-0.5cm}
 		\caption{Benign user account}
 		\label{fig:non-comp-1}
 	\end{subfigure}
 	\begin{subfigure}{0.24\textwidth}
 		\includegraphics[width=0.9\linewidth,keepaspectratio]{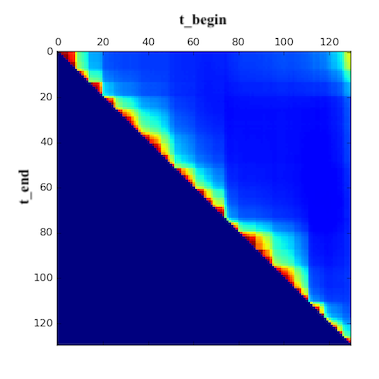}
 		\vspace{-0.5cm}
 		\caption{Benign user account}
 		\label{fig:non-comp-2}
 	\end{subfigure}
 	\begin{subfigure}{0.24\textwidth}
 		\includegraphics[width=0.9\linewidth,keepaspectratio]{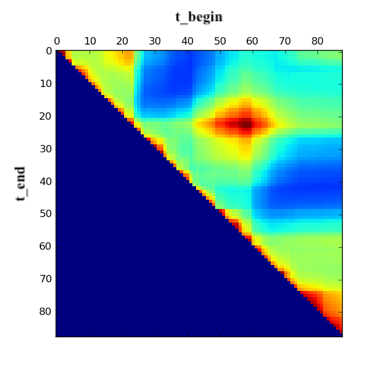}
 		\vspace{-0.5cm}
 		\caption{Compromised user account}
 		\label{fig:comp-1}
 	\end{subfigure}
 	\begin{subfigure}{0.24\textwidth}
 		\includegraphics[width=0.9\linewidth,keepaspectratio]{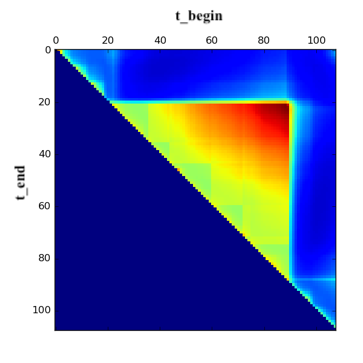}
 		\vspace{-0.5cm}
 		\caption{Compromised user account}
 		\label{fig:comp-2}
 	\end{subfigure}
 	
 	\caption{KL-divergence heatmap for different benign (Figure \ref{fig:non-comp-1} and \ref{fig:non-comp-2}) and compromised user accounts (Figure \ref{fig:comp-1} and \ref{fig:comp-2}). The x-axis of each figure depicts different values for $t_{begin}$, while the y-axis depicts different values for $t_{end}$. The color palette ranges from blue (low KL-divergence) to red (high KL-divergence).}
 	\label{fig:kl-div-heatmaps}
 \end{figure*}

\subsection{Language Model Similarity of Compromised Accounts}
\label{ssec:kl-compromised-accounts}

In the first part of our feasibility analysis we answer research question~\textbf{(\rom{1})} and study to what extend a KL-divergence measure can detect incoherence. 

To achieve this, we manually inspect the differences in user accounts by leveraging a heatmap as a visual cue. Figure \ref{fig:kl-div-heatmaps} depicts four heatmaps of two benign (Figures~\ref{fig:non-comp-1},~\ref{fig:non-comp-2}) and compromised user accounts (Figures~\ref{fig:comp-1},~\ref{fig:comp-2}). The x-axis of each figure depicts different values for $t_{begin}$, while the y-axis depicts different values for $t_{end}$. The color palette ranges from blue (low KL-divergence) to red (high KL-divergence). The left part of the plot below the diagonal is intentionally left empty, since these values would represent nonsensical variable assignments for $t_{begin}$ and $t_{end}$.

It is immediately evident that we find more high KL-divergence values for compromised compared to benign accounts. In the figures, high values are expressed by red and dark red colors. From this observation it can be inferred that the average KL-divergence will be higher for accounts that are compromised. By manually inspecting over 100 of these user account heatmaps we find that many of them follow this general trend. We understand this as preliminary evidence that a method which utilizes KL-divergence for detecting compromised accounts is feasible.

We further noticed in Figures \ref{fig:comp-1} and \ref{fig:comp-2} that the account takeover happened where the KL-divergence reaches its maximum (see the dark circle in Figure \ref{fig:comp-1} and tip of the pyramid in Figure \ref{fig:comp-2}). Unfortunately, this is not true for all inspected accounts, but it gives reason to believe that there might be potential to find the most likely period of an account takeover with our current framework. This could be done by finding $t_{begin}$ and $t_{end}$ that maximizes the difference of the language models $\theta^{User}$ and $\theta^{Attack}$.

\subsection{Estimate KL-divergence Using Random Sampling}
\label{ssec:kl-random-sampling}

In our second feasibility analysis we investigate research question~\textbf{(\rom{2})}, whether the average KL-divergence of a user account can be approximated using random sampling. More specifically, we try to find a reasonable estimate by calculating the KL-divergences only for a subset of the $t_{begin}$ and $t_{end}$ pairs.

\begin{figure}
    \centering
	\includegraphics[width=0.70\linewidth,keepaspectratio]{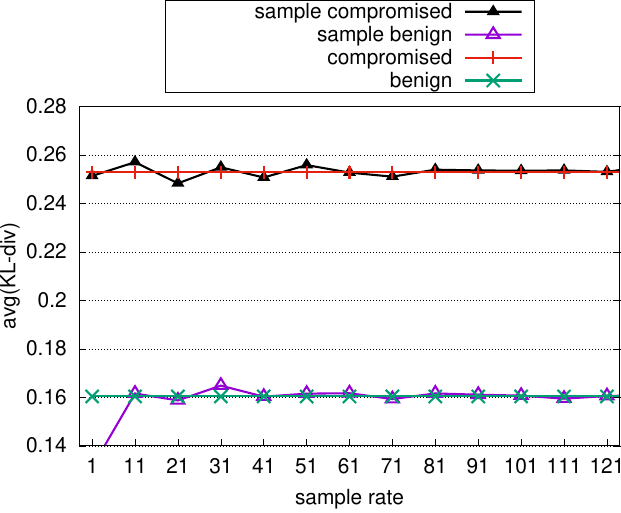}
	\caption{Average KL-divergence for different sample rates.}
	\label{fig:est-kl-div}
\end{figure}

Since we have calculated the KL-divergence for every possible combination of  $t_{begin}$ and $t_{end}$ for all of the 495 users, we know the actual average KL-divergence. We then try different sampling rates. For every sampling rate we average over the samples and compare them to the actual average that is calculated using all $t_{begin}$, $t_{end}$ pairs. 
The result is shown in Figure \ref{fig:est-kl-div}. In the figure we plot the actual average KL-divergence for compromised and benign against the averaged samples for different sample rates. The sample rates range from 1 to 121. The plot shows that the average KL-divergence for compromised accounts is about 0.1 higher than for benign accounts. This confirms our findings in Section~\ref{ssec:kl-compromised-accounts}. Furthermore we see that for small sample rates ($<81$) there are minimal deviations for the average ($\pm 0.01$). The higher the sample rate the lower these deviations become, as our estimate gets better. 

\begin{figure}
    \centering
	\includegraphics[width=0.65\linewidth,keepaspectratio]{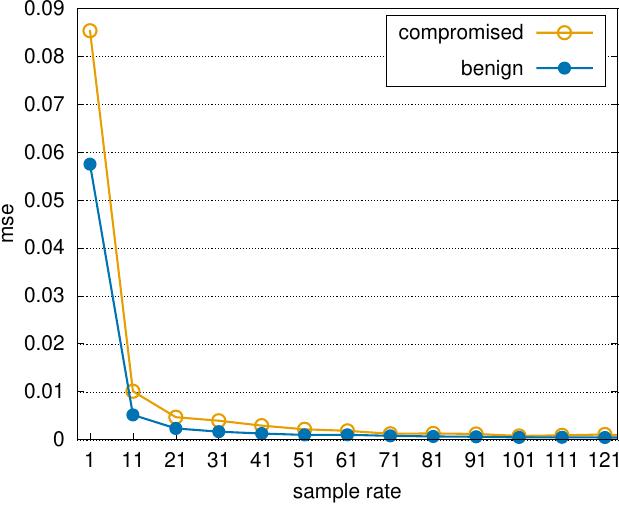}
	\caption{Mean squared error (mse) for different sample rates.}
	\label{fig:est-mse}
\end{figure}

Since our estimates only deviate slightly we also investigate the mean squared error (\textit{mse}). Here, the mse is defined as: $\frac{1}{n} \sum_{u \in U^{test}} (sampled\_avg(u) - actual\_ avg(u))^2$.
In Figure~\ref{fig:est-mse} we plot the mse for compromised and benign accounts. For very small sampling rates ($<50$) we see errors of over 0.07 and over 0.06 for compromised and benign accounts, respectively. Once the sample rate is greater than 101 the mse is close to 0. We therefore conclude that a sampling rate in the range of [50, 100] is sufficient for our experiments.  

\subsection{Effectiveness on Simulated Data}

In this subsection we show the effectiveness of our language model based method when detecting compromised accounts in a simulated environment. We answer research question \textbf{(\rom{3})}, compare the performance to general state-of-the-art text classification features and \textbf{(\rom{4})}, show that improvements can be made when combining the language model features with other specialized approaches, proposed in previous work in compromised account detection.

We perform an ablation study in Table~\ref{tab:classification-performance}, which shows the performance of all features together and individually. In this experiment 50\% of the tweets of a compromised account are compromised. We observe that we gain maximum performance over all metrics when all features are utilized (see column ``All''). The classifier reaches an accuracy of 0.80, high precision (0.90) and Recall at 0.68. Both measures are combined in the $F_1$-score, which reaches 0.78. We would like to note that a precision of 0.90 can be sufficient for many practical applications, where the system could alert users or platform providers.
When features are investigated individually, we find that the features ``Max'' and ``Variance'' perform worst.
The features ``Min'' and ``Mean'' perform almost equally in isolation and much better than the other two. We find the minimum sample of the two probability distributions to be a strong signal. This further confirms our earlier observations that compromised accounts can be distinguished by their higher average KL-divergence.

\begin{table}[t]
	\centering
	\caption{Ablation study using different measures.}
	\label{tab:classification-performance}
	\begin{tabular}{c||c c c c c}
    	\hline
        Measure   & All      & Max  & Min  & Mean & Var. \\
		\hline \hline
		Accuracy  & \textbf{0.80} & 0.59 & 0.76 & 0.75 & 0.48  \\
		$F_1$     & \textbf{0.78} & 0.57 & 0.72 & 0.72 & 0.35 \\
		Precision & \textbf{0.90} & 0.61 & 0.87 & 0.83 & 0.47 \\
		Recall    & \textbf{0.68} & 0.53 & 0.61 & 0.64 & 0.27 \\
		\hline
	\end{tabular}
\end{table}

We turn to research question~\textbf{(\rom{3})} and investigate how our language model based features compares to state-of-the-art text classification features. Table~\ref{tab:classifier-accuracy} lists the results for our features (\featurelm{}) and \featuredtov{}. Best performance is reached when 50\% of tweets are compromised, with accuracy at 0.80 and 0.71, respectively. When the number of compromised tweets is reduced by half, the accuracy drops to 0.75 and 0.69. When the amount of compromised tweets is set to 10\% and 5\%, accuracy further reduces for both features. For random ratios, the performance is slightly lower than for the fixed 25\% ratio. It is expected that this performance falls somewhere within the lower (5\%) and upper (50\%) bounds for performance. Summarizing, we see that \featurelm{} achieves higher Accuracy and Precision, whereas \featuredtov{} achieves higher Recall.

\begin{table*}[t]
	\centering
	\caption{\featurelm{} and \featuredtov{} features, with different percentages of compromised tweets.}
	\label{tab:classifier-accuracy}
	\begin{tabular}{c||c c c c|c c c c}
	    \multicolumn{1}{c}{} & \multicolumn{4}{c}{\featurelm{}} & \multicolumn{4}{c}{\featuredtov{}} \\
		\%& Accuracy           & $F_{1}$         & Precision       & Recall & Accuracy & $F_{1}$ & Precision & Recall\\
		\hline \hline
		$50$ & $\mathbf{0.80}$ & $\mathbf{0.78}$ & $\mathbf{0.90}$ & $0.68$ & $0.71$          & $0.71$          & $0.72$ & $\mathbf{0.71}$\\
		$25$ & $\mathbf{0.75}$ & $\mathbf{0.71}$ & $\mathbf{0.82}$ & $0.63$ & $0.69$          & $0.69$          & $0.70$ & $\mathbf{0.69}$\\
		$10$ & $\mathbf{0.65}$ & $0.62$          & $\mathbf{0.69}$ & $0.56$ & $0.63$          & $\mathbf{0.63}$ & $0.64$ & $\mathbf{0.63}$\\
		$5$  & $\mathbf{0.59}$ & $0.56$          & $\mathbf{0.60}$ & $0.52$ & $\mathbf{0.59}$ & $\mathbf{0.59}$ & $0.59$ & $\mathbf{0.59}$\\
		RND  & $\mathbf{0.75}$ & $\mathbf{0.70}$ & $\mathbf{0.81}$ & $0.61$ & $0.68$          & $0.68$          & $0.69$ & $\mathbf{0.68}$\\
		\hline
	\end{tabular}
\end{table*}

\begin{table*}[t]
    \centering
    \caption{Accuracy for different features and their combinations. Our method (\featurelm{}) is compared to general text representations.}
    \label{tab:comp-features}
    \begin{tabular}{c|c|c|c|c|c|c|c|c}
    \hline
    \multicolumn{1}{ p{0.3cm}|}{\centering \% } &
    \multicolumn{1}{|p{1.2cm}|}{\centering \featurecount{}} &
    \multicolumn{1}{|p{1.2cm}|}{\centering \featuretfidf{}} &
    \multicolumn{1}{|p{1.2cm}|}{\centering \featuredtov{}} &
	\multicolumn{1}{|p{1.7cm}|}{\centering \featuredtov{} + \newline \featuretfidf{}} & 
    \multicolumn{1}{|p{1.2cm}|}{\centering \featurelm{}} &
    \multicolumn{1}{|p{1.2cm}|}{\centering \featurelm{} + \newline \featuretfidf{}} & 
    \multicolumn{1}{|p{1.2cm}|}{\centering \featurelm{} + \newline \featuredtov{}} & 
    \multicolumn{1}{|p{1.1cm}}{\centering all } \\
    \hline
    50 & 	 $0.53$	 & 	$0.56$ &	$0.71$ & 	$0.72$ &	$0.80$ & 	$0.81$ & 	\textbf{0.87} &	\textbf{0.87} \\
    25 &	 $0.53$	 & 	$0.55$ &	$0.69$ & 	$0.69$ &	$0.75$ & 	$0.75$ & 	\textbf{0.82} &	\textbf{0.82} \\
    10 &	 $0.52$	 & 	$0.54$ &	$0.63$ & 	$0.63$ &	$0.65$ & 	$0.65$ & 	0.70 &	\textbf{0.71} \\
    5  &	 $0.52$	 & 	$0.53$ &	$0.59$ & 	$0.59$ &	$0.59$ & 	$0.59$ & 	\textbf{0.62} &	\textbf{0.62} \\
    RND&	 $0.53$	 & 	$0.55$ &	$0.68$ & 	$0.68$ &	$0.74$ & 	$0.74$ & 	\textbf{0.80} &	\textbf{0.80} \\
    \hline
    \end{tabular}
\end{table*}

In Table \ref{tab:comp-features} we compare \featurelm{} to general text classification features and their combinations. It can be seen that \featuredtov{} performs better than the word based models.  This might be due to the fact that \featuredtov{} is able to learn a better representation for each document, compared to \featuretfidf{} and \featurecount{}. \featuredtov{} outperforms the \featuretfidf{} model by up to 15 percentage points ($\approx$27\% relative improvement), if the amount of compromised tweets reaches 50\%.  The performance improvement is less drastic when only small amounts of tweets are compromised. 
We further find that \featurelm{} performs best, 
as it outperforms \featuretfidf{} and \featuredtov{} with 6 and 7 percentage points when only 5 percent of tweets are compromised, respectively. The most distinguishing performance is achieved when the amount of compromised tweets reaches 50\%. There, \featurelm{} outperforms \featuredtov{} with 9 percentage points ($\approx$13\% relative improvement) and \featuretfidf{} with 24 percentage points ($\approx$43\% relative improvement). We argue that this is strong evidence for the superiority of the \featurelm{} feature compared to other general text classification features.

When combining features, we find that adding \featuredtov{} features to \featurelm{} results in the highest performance improvements, with up to 7 percentage points over \featurelm{} alone (column: ``\featurelm{} + \featuredtov{}''). Adding \featuretfidf{} to \featurelm{} or \featuredtov{} does only have a negligible effect (columns: ``\featurelm{} + \featuretfidf{}'' and ``\featuredtov{} + \featuretfidf{}''). The same minimal improvements are observed when adding \featuretfidf{} to \featurelm{} and \featuredtov{} (column: ``all''). Thus, we conclude that the proposed \featurelm{} features add meaningful signals to general text classification features.

\subsection{Comparison with Existing Detection Methods}

For answering research question~\textbf{(\rom{4})}, we show how \featurelm{} performs in comparison to other compromised account detection methods and that improvements can be made by combining \featurelm{} with these methods. The baseline features here differ in the way that they are specifically designed for compromised account detection, whereas the previous features were general textual representations. 

We compare all models in Table~\ref{tab:related-methods}. The first three rows compare the compromised account detection features on their own. Our model outperforms the strongest baseline (COMPA) over all metrics, with the highest gains made in accuracy and precision, increasing 19.4\% and 26.6\%, respectively. The fifth through seventh row show the combination of the \featurelm{} features with the two baselines alone and in combination. Here we find that our method can further improve when the baseline features are added. We argue that our features are orthogonal to existing work, as evidenced by the performance improvement due to combination of the feature spaces. The second-to-last row shows the performance, when all baseline features are combined with the \featurelm{} model. This results in the best performing model, with gains over all metrics. 

\begin{table*}[t]
	\centering
	\caption{Comparison to features from related methods. The dataset has random percentages of tweets compromised (RND).}
	\label{tab:related-methods}
	\begin{tabular}{l|c|c|c|c}
		\hline
		Model              & Accuracy           & $F_{1}$         & Precision       & Recall \\
		\hline \hline
		COMPA \cite{DBLP:conf/ndss/EgeleSKV13}      & $0.62$ & $0.60$ & $0.64$ & $0.56$ \\
		VanDam \cite{vandam2017understanding}       & $0.50$ & $0.47$ & $0.50$ & $0.45$ \\
		\featurelm{}                                & \textbf{0.74} & \textbf{0.70} & \textbf{0.81} & \textbf{0.61} \\
		\hline
		\quad improvement \featurelm{} over best baseline
		                                            & 19.4\% & 16.7\% & 26.6\% & 8.9\%  \\
		\hline
		\featurelm{} + COMPA                        & $0.75$ & \textbf{0.73} & $0.81$ & $0.66$ \\
		\featurelm{} + VanDam                       & $0.74$ & $0.71$ & \textbf{0.82} & $0.62$ \\
		\featurelm{} + COMPA + VanDam               & \textbf{0.76} & \textbf{0.73} & $0.81$ & \textbf{0.67} \\ 
        \hline
        \quad improvement over \featurelm{}               & 2.7\% & 4.3\% & 1.2\% & 9.8\% \\
		\hline
		\featurelm{} + \featuredtov{} + \featuretfidf{} + COMPA + VanDam 
		                                            & \textbf{0.81}  & \textbf{0.79}  & \textbf{0.85} & \textbf{0.75} \\
		\hline
		\quad improvement when adding standard features   & 6.6\% & 8.2\% & 4.9\%& 11.9\% \\
		\hline
	\end{tabular}
\end{table*}

\subsection{Effectiveness on Real Data}

The final question we answer qualitatively is \textbf{(\rom{5})}, how effective is our method on non-simulated data. We sort 20 accounts with the highest probability of being compromised into six categories, shown on the left in Table \ref{tab:category-assignment}. We find that most of these accounts belong to categories with high variation in language, i.e., news, spam, re-tweet bot. One of the accounts is found to be compromised. This result needs to be seen in perspective, since the classifier was trained on simulated data. We would expect the percentage of compromised accounts to be much higher, if training data with ``real'' labels are used.
These results show that our algorithm can also detect ``unusual'' accounts and users, thus potentially enabling development of novel text mining algorithms for analyzing user behavior on social media, which should be an interesting future research topic. 

\begin{table}[t]
	\centering
	\caption{Statistics of manually evaluated accounts.}
	\label{tab:category-assignment}
	\begin{tabular}{c|c||c|c}
		\hline
		Category & Count & Status & Count \\
		\hline \hline
		News & $5$ & Abandoned & $7$\\
		Spam & $4$ & Active & 6\\
		Re-tweet Bot& $2$ & Deleted & 4 \\
		Compromised  & $1$ & Protected & 2 \\
        Regular & $7$ & Suspended & 1 \\
        Unknown & $1$ & &\\
		\hline
	\end{tabular}
\end{table}

We further inspect the current state of each account on the right side of Table~\ref{tab:category-assignment}. We find that most accounts are abandoned and one account was suspended by Twitter. The account that was identified as compromised was set to protected, which could be an indicator that this user had become more conscious about tweets that were posted from her account and therefore decided to not share her tweets publicly.   
While manually investigating the account's tweets, we find that after discussing general topics a near-duplicate message is posted hundreds of times with only brief pauses between tweets. Different users were addressed directly, which were most likely followers of the account. With this scheme the attacker tries to directly grab the attention of a targeted user. The messages included one of two links that were identified as suspicious by the link-shortening service the hacker utilized to hide the actual URL.
After the attack, the tweets return to discuss similar topics as before. 
From the content of this messages we conclude that the hacker was pursuing a led generation scheme \cite{DBLP:conf/ccs/ThomasLGP14}, where users are lured into clicking a link. 
It is reasonable to assume that if our algorithm were applied at much larger scale to all the Twitter users, it would most likely be able to detect many more compromised accounts. 

\section{Conclusion and Future Work}
\label{sec:future-work}

We proposed a novel general framework based on semantic text analysis for detecting compromised social media accounts. Following the framework, we proposed a specific instantiation based on uni-gram language models and KL-divergence measure, and designed features accordingly for use in a classifier
that can distinguish compromised from benign accounts.
We conclude that (1) the proposed \featurelm{} feature is most effective, even when used as a single feature-based detection method. (2) \featurelm{} captures new signals that haven't been captured in the existing methods and features, which is shown by the further improvement when added on top of the baselines. (3) The best performing method would combine the proposed LM with all the existing features. 
Although \featurelm{} is motivated by a security problem, our general idea of performing differential semantic analysis of text data may be applicable to other domains where incohesion (or outlier) in text data needs to be captured. Further exploration of such potential directions is interesting future work.

\bibliographystyle{IEEEtranN}
\bibliography{IEEEabrv,main}

\end{document}